\documentstyle[12pt,fleqn]{article}
\def\singlespace {\smallskipamount=3.75pt plus1pt minus1pt
                  \medskipamount=7.5pt plus2pt minus2pt
                  \bigskipamount=15pt plus4pt minus4pt
                  \normalbaselineskip=15pt plus0pt minus0pt
                  \normallineskip=1pt
                  \normallineskiplimit=0pt
                  \jot=3.75pt
                  {\def\smallskip {\vskip\smallskipamount}}
                  {\def\medskip   {\vskip\medskipamount}}
                  {\def\bigskip   {\vskip\bigskipamount}}
                  {\setbox\strutbox=\hbox{\vrule 
                    height10.5pt depth4.5pt width 0pt}}
                  \parskip 7.5pt
                  \normalbaselines}
\def\middlespace {\smallskipamount=5.625pt plus1.5pt minus1.5pt
                  \medskipamount=11.25pt plus3pt minus3pt
                  \bigskipamount=22.5pt plus6pt minus6pt
                  \normalbaselineskip=22.5pt plus0pt minus0pt
                  \normallineskip=1pt
                  \normallineskiplimit=0pt
                  \jot=5.625pt
                  {\def\smallskip {\vskip\smallskipamount}}
                  {\def\medskip   {\vskip\medskipamount}}
                  {\def\bigskip   {\vskip\bigskipamount}}
                  {\setbox\strutbox=\hbox{\vrule 
                    height15.75pt depth6.75pt width 0pt}}
                  \parskip 11.25pt
                  \normalbaselines}
\def\doublespace {\smallskipamount=7.5pt plus2pt minus2pt
                  \medskipamount=15pt plus4pt minus4pt
                  \bigskipamount=30pt plus8pt minus8pt
                  \normalbaselineskip=30pt plus0pt minus0pt
                  \normallineskip=2pt
                  \normallineskiplimit=0pt
                  \jot=7.5pt
                  {\def\smallskip {\vskip\smallskipamount}}
                  {\def\medskip   {\vskip\medskipamount}}
                  {\def\bigskip   {\vskip\bigskipamount}}
                  {\setbox\strutbox=\hbox{\vrule 
                    height21.0pt depth9.0pt width 0pt}}
                  \parskip 15.0pt
                  \normalbaselines}

\include{dspace12}
\def\be{\begin{equation}}
\def\ee{\end{equation}}
\def\bea{\begin{eqnarray}}
\def\eea{\end{eqnarray}}
\def\nn{\nonumber}
\def\th{\theta}
\def\ph{\phi}
\def\lt{\left}
\def\rt{\right}
\def\sect #1{\setcounter{equation}{0}}

\begin{document}
\singlespace
\begin{flushright}
{\bf hep-th/9406148}
\end{flushright}
\vspace{0.5in}
\begin{center}
{\LARGE {A radiating dyon solution }}
\end{center}
\vspace{0.8in}
\begin{center}
{\large{A. Chamorro\footnote[1]{E-mail\ :\ wtpchbea@lg.ehu.es} 
  and K. S. Virbhadra\footnote[2]{
Present address\ :\ Theoretical Astrophysics Group, Tata Institute
of Fundamental Research, Homi Bhabha Road, Colaba, Bombay
400005,  India; E-mail\ :\  shwetketu@tifrvax.tifr.res.in}\\
Departamento de F\'{\i}sica Te\'{o}rica\\
Universidad del Pa\'{\i}s Vasco\\
Apartado 644, 48080 Bilbao, Spain\\
}}
\end{center}
\vspace{0.7in}
\begin{abstract}
We give a non-static exact solution of the Einstein-Maxwell equations (with 
null fluid), which is a non-static magnetic charge generalization to the
Bonnor-Vaidya solution and 
describes the gravitational and electromagnetic fields of a nonrotating massive
radiating dyon. In addition, using the energy-momentum pseudotensors of 
Einstein  and Landau and Lifshitz we obtain the energy, momentum, and power 
output of the radiating dyon and find that both prescriptions give the same 
result.
\end{abstract}
\begin{flushleft}
\end{flushleft}
\vspace{0.7in}
\begin{center}
 Pramana - J. Phys. {\bf 45} 181 1995
\end{center}
\newpage
There has been considerable interest in obtaining non-static solutions of
Einstein's equations describing the gravitational field of a star radiating
null fluid $[1-3]$. In $1953$, Vaidya $[2]$ obtained a nice form of non-
static generalization to the Schwarzschild solution and it beacame well
 known
in the literature after the discovery of quasars. 
The quasars are high energy sources and therefore their gravitational field
cannot be described by the Schwarzschild metric while the Vaidya metric
is relevant to the study of such objects.
Vaidya and others gave non-static  generalizations to the Kerr solution $[3]$.
Bonnor and Vaidya $[4]$ obtained a 
non-static generalization to the   Reissner-Nordstr\"{o}m (RN) solution
describing the emission of charged null fluid from a spherically symmetric
charged radiating body. 
Mallett $[5]$ gave an exact solution describing the radiating Vaidya metric $[1]$ 
in a de Sitter universe. Patino and Rago $[6]$ obtained a solution of
the  Einstein-Maxwell (EM) equations with null fluid for a spherically symmetric  
radiating massive charged (electric) object in a de Sitter universe.
 
The existence of magnetic monopoles is not yet confirmed, but it has been 
a subject of interest of many physicists (see in ref. $7$). The Bonnor-Vaidya
solution $[4]$  is not enriched  with  a magnetic charge parameter. Therefore,
 it is of interest to obtain  a non-static magnetic charge generalization to the 
Bonnor- Vaidya solution,  characterized by three time-dependent parameters:
mass, electric charge, and magnetic charge. 
Further we calculate  the energy, momentum and power output of the radiating
dyon in prescriptions of Einstein as well as Landau and Lifshitz (LL).
Throughout this  paper we use  geometrized units where the
 gravitational constant $G\ =\ 1$ and the speed of light in vacuum $c\ =\ 1$.
We follow the convention that   Latin indices take values from $0$ to $3$
($x^0$ is the time coordinate) and  Greek indices take values from $1$ to $ 3$.

The EM equations with null fluid present are $[4]$
\be
R_i^{\ k} - \frac{1}{2}\ g_i^{\ k} R = 8 \pi \lt( E_i^{\ k} +
N_i^{\ k}\rt)\ ,
\ee
\be
\frac{1}{\sqrt{-g}} \lt(\sqrt{-g}\ F^{ik}\rt)_{,k} = 4 \pi J_{(e)}^i\ ,
\ee
\be
\frac{1}{\sqrt{-g}} \lt(\sqrt{-g}\ {}^\star F^{ik}\rt)_{,k} = 4 \pi J_{(m)}^i\ ,
\ee
where
\be
E_i^{\ k} = \frac{1}{4\pi} \lt[-F_{im} F^{km} + \frac{1}{4}\  
                              g_i^{\ k} F_{mn} F^{mn}\rt]
\ee
is the energy-momentum tensor of the electromagnetic field
and
\be
N_i^{\ k} = V_i \ V^k\ 
\ee
is the energy-momentum tensor of the null fluid.
$V^k$ is the null fluid current vector satisfying
\be
g_{ik}\ V^i \ V^k \ = \ 0 \ .
\ee

${{}^{\star}F}^{ik}$, the dual of the electromagnetic
field tensor $F^{ik}$,  is given by
\be
{{}^{\star}F}^{ik} = \frac{1}{2 \sqrt{-g}} \varepsilon^{iklm} \ F_{lm}\ .
\ee
$\varepsilon^{iklm}$ is the Levi-Civita tensor density.
$R_i^{\ k}$ is the Ricci tensor.
 $J_{(e)}^i$  $(J_{(m)}^i)$ stands
for the electric (magnetic) current density vector.

An exact solution of the above equations
describing the gravitational and electromagnetic fields of a non-rotating
radiating dyon is given, in coordinates
$x^0 = u, x^1 = r, x^2 = \th, x^3 = \phi $,  by the metric,
\be
ds^2\ =\ B\ du^2\ +\ 2 \ du\ dr\ - \ r^2 \lt(d\th^2 + sin^2\th\ d\ph^2\rt)\ ,
\ee
where
\be
B\ = \ 1\ -\ \frac{2 M \lt(u\rt)}{r}\ +\ \frac{{q_e\lt(u\rt)}^2  + 
 {q_m\lt(u\rt)}^2}{r^2} 
\ee
and the non-zero components of the electromagnetic field tensor ,
\bea
F_{10}\ &=& \ \frac{q_e\lt(u\rt)}{r^2}\ , \nn\\
F_{23}\ &=& \ q_m(u)\ sin\th \ .
\eea
$M(u), q_e(u),$ and $q_m(u)$ are  mass, electric and magnetic 
charge parameters, respectively. These parameters depend on the retarded time
coordinate $u$.

The surviving components of the Einstein tensor, $G_i^{\ k} \equiv R_i^{\ k}
- \frac{1}{2}\ g_i^{\ k}\ R$\ ,  the energy-momentum tensor of the 
electromagnetic field, $E_i^{\ k}$, and the energy-momentum tensor of the
null fluid, $N_i^{\ k}$,   are
\bea
G_0^{\ 0}\ &=&\ G_1^{\ 1}\ =\ -\ G_2^{\ 2}\ =\ -\ G_3^{\ 3}\ = \ 
\frac{{q_e}^2  +   {q_m}^2}{r^4}\ ,\nn\\
G_0^{\ 1} &=& \ K^2\ ,
\eea

\be
E_0^{\ 0}\ =\ E_1^{\ 1}\ =\ -\ E_2^{\ 2}\ =\ -\ E_3^{\ 3}\ = \ 
\frac{{q_e}^2  +   {q_m}^2}{8 \pi r^4}\ ,
\ee
\be
N_0^{\ 1}\ = \ \frac{K^2}{8 \pi}\ ,
\ee
where
\be
K^2\ = \ \frac{2 \lt( q_e  \dot{q_e} + q_m \dot{q_m}  - \dot{M} r \rt)}{r^3}\ .
\ee
The dot denotes the derivative with respect to the retarded time coordinate $u$.
The null fluid current vector, the electric and magnetic current density vectors
are 
\be
V^i\ =\ g_1^{\ i}\ \frac{K}{\sqrt{8 \pi}}\ ,
\ee
\bea
J_{(e)}^{\ i}\ =\ - \ \frac{\dot{q_e}}{4\ \pi \ r^2}\ g_1^{\ i} \ ,\nn\\
J_{(m)}^{\ i}\ =\ - \ \frac{\dot{q_m}}{4\ \pi \ r^2}\ g_1^{\ i} \ .
\eea
The radiating dyon solution given by us yields :
$(a)$\ the Bonnor-Vaidya solution  when $q_m = 0$
$(b)$\ the Vaidya radiating star solution $[2]$ when $q_e = q_m = 0$
$(c)$\ the Reissner-Nordstr\"{o}m solution when $q_m = 0 and q_e$ and $M$ are
      constants
$(d)$\ the Schwarzschild solution when $q_e = q_m = 0$ and $M$ is constant
and
$(e)$\ the static dyon solution $[8]$ when $M, q_e,$ and $q_m$ are constants.

Using the Tolman pseudotensor, 
Vaidya $[9]$ calculated the total energy of a spherically symmetric radiating
star  and got $E = M$. Further Lindquist, Schwartz and Misner (LSM)$[10]$,
 using the LL  pseudotensor, obtained  the energy, momentum, and power output
 for the Vaidya spacetime and  found  that the total energy and momentum
components are  $p^i = M;0,0,0$ and the power output is  $-dM/du$.
One of the present authors $[11]$, using several energy-momentum pseudotensors,
calculated the energy and momentum components for the Vaidya metric 
and found  the same result as obtained by LSM. Now we  obtain the energy, 
momentum, and power output for the radiating dyon in prescriptions of Einstein 
as well as LL and show that both give the same result. The  energy-momentum 
pseudotensors of Einstein $[12]$  and LL  $[13]$ are 
\be
\Theta_i^{\ k} = \frac{1}{16 \pi} \ {H_i^{\ kl}}_{,l}\ ,
\ee
where
\be
{H_i^{\ kl}}\  =\ - \ {H_i^{\ lk}}\ = \ 
\frac{g_{in}}{\sqrt{-g}}
\lt[ -g \lt( g^{kn} g^{lm} - g^{ln} g^{km} \rt) \rt]_{,m} \ \ ,
\ee
 and

\be
L^{mn} = L^{nm}  = \frac{1}{16 \pi} \ {S^{mjnk}}_{,jk}\ \ ,
\ee
where
\be
S^{mjnk}\ =\ 
 -g \left( g^{mn} g^{jk} - g^{mk} g^{jn}\right)\ .
\ee
They satisfy the local conservation laws:
\be
\frac{\partial \Theta_i^{\ k}}{\partial  x^k}\ =\ 0 \ ,
\ee
\be
\frac{\partial L^{mn}}{\partial x^m}\ =\ 0 \ .
\ee
$\Theta_0^{\ 0}$ is the energy density. $\Theta_{\alpha}^{\ 0}$ and 
$\Theta_0^{\alpha}$ are the momentum and energy current density components,
respectively.
$L^{00}$ and  $L^{\alpha 0}$ give, respectively, the  energy density 
and momentum (energy current) density components in LL prescription.
The energy and momentum components in the prescription of Einstein
are
\be
P_i\ =\ \int\int\int \Theta_i^{\ 0}\ dx^1 \ dx^2 \ dx^3\ ,
\ee
whereas  in the LL prescription are
\be
P^i\ =\ \int\int\int L^{0i}\ dx^1 \ dx^2 \ dx^3 \ .
\ee
$i=0$ gives the energy and $i=1,2,3$ give the momentum components.
 One knows  that the 
energy-momentum pseudotensors  give the correct result if calculations are 
carried out in quasi- cartesian coordinates $[9-14]$ . The quasi-cartesian coordinates
 are those in which the metric $g_{ik}$ approaches the Minkowski metric
$\eta_{ik}$ at great distances. Therefore, one transforms the line element $(8)$
, given in $u, r, \theta, \phi$ coordinates, to  quasi-cartesian coordinates 
$ t, x, y, z$ according to 
\bea
u\ &=&\ t\ -\ r \ ,\nn\\
x\ &=&\ r\ \sin\th\ \cos\ph \ , \nn\\
y\ &=&\ r\ \sin\th\ \sin\ph \ , \nn\\
z\ &=&\ r\ \cos\th \ ,
\eea
and gets 
\be
ds^{2} = dt^2 -dx^2-dy^2-dz^2-\lt(\frac{2M\lt(u\rt)}{r}-\frac{q_e\lt(u\rt)^2
+q_m\lt(u\rt)^2}{r^2}\rt)\
\lt[dt-\frac{x dx + y dy + z dz}{r}\rt]^2 \ ,
\ee
where, 
\be
r\ =\ \lt(x^2+y^2+z^2\rt)^{1/2} \ .
\ee
We are interested in calculating the energy-momentum components and the power 
output of the radiating dyon. Therefore, the required components of
$H_i^{\ kl}$ are
\bea
H_0^{\ 01}\ &=&\ \frac{2 x \Lambda}{r^4} \ , \nn\\
H_0^{\ 02}\ &=&\ \frac{2 y \Lambda}{r^4} \ , \nn\\
H_0^{\ 03}\ &=&\ \frac{2 z \Lambda}{r^4} \ , \nn\\
H_1^{\ 01}\ &=&\ \frac{ x^2 \Omega -Q^2 r^2}{r^5} \ , \nn\\
H_2^{\ 02}\ &=&\ \frac{ y^2 \Omega -Q^2 r^2}{r^5} \ , \nn\\
H_3^{\ 03}\ &=&\ \frac{ z^2 \Omega -Q^2 r^2}{r^5} \ , \nn\\
H_1^{\ 02}\ &=&\ H_2^{\ 01}\ =\ \frac{x y \Omega}{r^5} \ , \nn\\
H_2^{\ 03}\ &=&\ H_3^{\ 02}\ =\ \frac{y z \Omega}{r^5} \ , \nn\\
H_3^{\ 01}\ &=&\ H_1^{\ 03}\ =\ \frac{z x \Omega}{r^5} \ , \nn\\
H_0^{\ 12}\ &=&\ H_0^{\ 23}\ =\ H_0^{\ 31}\ = \ 0 \ ,
\eea
and those of $S^{mjnk}$ are
\bea
S^{0101}\ &=& \ \frac{\Lambda (x^2 - r^2) - r^4}{r^4} \ , \nn\\
S^{0202}\ &=& \ \frac{\Lambda (y^2 - r^2) - r^4}{r^4} \ , \nn\\
S^{0303}\ &=& \ \frac{\Lambda (z^2 - r^2) - r^4}{r^4} \ , \nn\\
S^{0102}\ &=& \ \frac{\Lambda x y}{r^4} \ , \nn\\
S^{0203}\ &=& \ \frac{\Lambda y z}{r^4} \ , \nn\\
S^{0301}\ &=& \ \frac{\Lambda z x}{r^4} \ , \nn\\
S^{0221}\ &=& \ S^{0331}\ = \ \frac{\Lambda x}{r^3} \ , \nn\\
S^{0112}\ &=& \ S^{0332}\ = \ \frac{\Lambda y}{r^3} \ , \nn\\
S^{0113}\ &=& \ S^{0223}\ = \ \frac{\Lambda z}{r^3} \ , \nn\\
S^{0123}\ &=& \ S^{0231}\ = \ S^{0312} = 0 \ ,
\eea
where
\bea
\Lambda\ &=&\ 2  M r - Q^2 \ , \nn\\
\Omega \ &=&\  3 Q^2\ - \ 4 M r \ , \nn\\
Q^2 \ &=& \ {q_e}^2 + {q_m}^2 \ .
\eea
We use $(17)$ and $(28)$ in $(23)$, and $(19)$ and $(29)$ in $(24)$,
apply the Gauss theorem and evaluate the integrals over the surface of 
two-sphere of radius $r_o$.  The energy and momentum components 
in both prescriptions (Einstein as well as LL) are
\be
E(r_0)\ = \ M - \frac{{q_e}^2 + {q_m}^2}{2 r_0}
\ee
and
\be
P_x\ =\ P_y\ =\ P_z\ = 0 \ .
\ee
Thus the total energy and momentum of the radiating dyon is
$P^i\ =\ M; 0,0,0$ as expected. Now using $(17)-(20)$ and $(28)-(30)$ we
calculate the energy current density components in both  prescriptions
and get
\be
\Theta_0^{\ 1}\ =\  L^{01} = x \Delta \ ,
\ee
\be
\Theta_0^{\ 2}\ =\  L^{02} = y \Delta \ ,
\ee
\be
\Theta_0^{\ 3}\ =\  L^{03} = z \Delta \ ,
\ee
where
\be
\Delta \  =\ \frac{ q_e \dot{q_e} + q_m \dot{q_m}- r \dot{M}}
            {4 \pi r^4} \ .
\ee
Again we get the same result in both  prescriptions.
Switching off the electric and magnetic charge parameters we  get the result
 for the Vaidya metric which were earlier obtained by one of the
 present authors $[11]$. 
Using the energy current density components, given by $(33-36)$, we calculate
the  power  output across a 2-sphere of radius $r_0$ and  get,
\be
W\lt(r_0\rt)\ =\ - \ \frac{d}{du}
\lt[   M \  -\ \frac{ {q_e}^2 +  {q_m}^2  }{2 r_0}    \rt] \ .
\ee

$q_e = q_m = 0$ in $(37)$ gives  the power output 
for the Vaidya spacetime $[10]$.  The total power output ( $r_0$
approaching infinity in  $(37)$ ) is $-dM/du$.
Without using any energy-momentum pseudotensor,
Bonnor and Vaidya $[3]$ obtained the power output for the metric given by them.
They  got the same result as  obtained by us.

\begin{flushleft}
{\bf Acknowledgements}
\end{flushleft}
This work has been partially supported by the Universidad del Pais 
Vasco under contract UPV 172.310 - EA062/93 ( AC ) and by a Basque 
Government post-doctoral fellowship (KSV).

This work was presented in the seventh Marcel Grossmann Meeting (1994)
and is to appear in the Proceedings in brief.
\newpage
\begin{flushleft}
{\bf References}

$[1]$  P. C. Vaidya, {\it Proc. Indian Acad. Sci.} {\bf A33}  264 (1951);\\
\ \ \ \ \ \ A. K. Raychaudhuri, {\it Z. Physik} {\bf 135}  225 (1953);\\
\ \ \ \ \ \  W. Israel, {\it Proc. Roy. Soc. (London)} {\bf A248}  404 (1958).\\
$[2]$  P. C. Vaidya, {\it Nature} {\bf 171}  260 (1953).\\
$[3]$ P. C. Vaidya and L. K. Patel, {\it Phys. Rev.} {\bf D7}  3590 (1973);\\ 
\ \ \ \ \ \ P. C. Vaidya, {\it Proc. Camb. Phil. Soc.} {\bf 75} 383 (1974);\\
\ \ \ \ \ \ P. C. Vaidya, L. K. Patel and P. V. Bhatt, {\it Gen. Rel. Grav.}
           {\bf 16} 355 (1976);\\
\ \ \ \ \ \ M. Carmeli and M. Kaye, {\it Ann. Phys. (N.Y.)} {\bf 103}  97 (1977).\\
$[4]$  W. B. Bonnor and P. C. Vaidya, {\it Gen. Rel. Grav.} {\bf 1}  127 (1970).\\
$[5]$  R. L. Mallett, {\it Phys. Rev.} {\bf D31}  416 (1985).\\
$[6]$  A. Patino and H. Rago, {\it Phys. Lett.} {\bf A121}  329 (1987).\\
$[7]$ A. S. Goldhaber and W. P. Trower, {\it Am. J. Phys.} {\bf 58}  429
 (1990).\\
$[8]$ I. Semiz, {\it Phys. Rev.} {\bf D46} 5414 (1992).\\
$[9]$ P. C. Vaidya, {\it J. Univ. Bombay} {\bf 21}  1 (1952). \\
$[10]$  R. W. Lindquist, R. A. Schwartz, and C. W. Misner,
{\it Phys. Rev.} {\bf 137} B1364 (1965).\\
$[11]$  K. S. Virbhadra, {\it Pramana-J. Phys.} {\bf 38}  31 (1992). \\
$[12]$  C. M\o ller, {\it Ann. Phys. (NY)} {\bf 4}  347 (1958). \\
$[13]$  L. D. Landau and E. M. Lifshitz,  {\it The Classical Theory of Fields} (
         Pergamon Press,\\
\ \ \ \ \ \ \ Oxford, 1985) p.280.\\
$[14]$  K. S. Virbhadra, {\it Phys. Rev.} {\bf D41}  1086 (1990);\\
\ \ \ \ \  K. S. Virbhadra, {\it Phys. Rev.} {\bf D42}  2919 (1990);\\
\ \ \ \ \  F. I. Cooperstock and S. A. Richardson, in {\it Proc. 4th Canadian Conf.
on General\\
\ \ \ \ \ \ \ Relativity and Relativistic Astrophysics} ( World Scientific, 
Singapore, 1991 );\\
\ \ \ \ \  K. S. Virbhadra and J. C. Parikh, {\it Phys. Lett.} {\bf B317} 
 312 (1993);\\
\ \ \ \ \  K. S. Virbhadra and J. C. Parikh, {\it Phys. Lett.} {\bf B331} 
 302 (1994).\\
\end{flushleft}

\end{document}